\documentclass[a4paper,11pt]{article}
\pdfoutput=1 

\usepackage{jcappub} 

\usepackage[T1]{fontenc} 
\usepackage{mathtools}
\usepackage{siunitx}

\usepackage{enumitem}
\usepackage{siunitx}
\usepackage[utf8x]{inputenc}

\usepackage{aas_macros}
\usepackage{amssymb}
\usepackage{cleveref}
\usepackage{siunitx}
\usepackage{enumerate}
\usepackage[dvipsnames]{xcolor}
\usepackage{amsmath}

\usepackage{graphicx} 
\usepackage{array}

\usepackage{amsthm}

\usepackage{psfrag}
\usepackage{color}
\usepackage{graphicx}
\usepackage{upgreek}
\usepackage{feynmp}
\DeclareMathAlphabet{\mathpzc}{OT1}{pzc}{m}{it}

\usepackage{tikz}
\usetikzlibrary{trees}
\usetikzlibrary{decorations.pathmorphing}
\usetikzlibrary{decorations.markings}
\tikzset{
    photon/.style={decorate, line width=0.15mm, decoration={snake,amplitude=3pt,segment length=8pt}, draw=black},
    wino/.style={draw=redwine},    
    fermion/.style={draw=black, line width=0.2mm, postaction={decorate},
        decoration={markings,mark=at position .55 with {\arrow[draw=black,scale=2,#1]{>}}}},
    scalar/.style={draw=black, dashed,postaction={decorate},
        decoration={markings,mark=at position .55 with {\arrow[draw=black,scale=2,#1]{>}}}},
    scalarline/.style={draw=black, postaction={decorate},
        decoration={markings,mark=at position .55 with {\arrow[draw=black,scale=2,#1]{>}}}},
    scalarline2/.style={draw=black, postaction={decorate} },
    scalar2/.style={draw=black, dashed,postaction={decorate}},
    gluon/.style={decorate, draw=black,
        decoration={coil,amplitude=3pt, segment length=4pt}},
    graviton/.style={decorate, draw=black,
        decoration={zigzag,amplitude=3pt, segment length=4pt}}
}
\tikzstyle{blob}=[circle,
                              minimum size=20,
                              draw=black!80,
                              fill=black!80]   
\tikzstyle{redblob}=[circle,
                              thick,
                              minimum size=0.4cm,
                              draw=red!80,
                              fill=red!60]

\definecolor{darkgreen}{rgb}{0,0.5,0}

\allowdisplaybreaks[4]

\definecolor{darkgreen}{rgb}{0,0.5,0}

\newcommand{\comment}[1]{}

\newcommand{\bseq}{\begin{subequations}}
\newcommand{\eseq}{\end{subequations}}
\newcommand{\be}{\begin{equation}}
\newcommand{\ee}{\end{equation}}

\newcommand{\RN}[1]{%
  \textup{\uppercase\expandafter{\romannumeral#1}}%
}

\renewcommand{\tanh}{\mathop{\rm th}\nolimits}

\renewcommand{\ln}{\mathop{\rm ln}\nolimits}

\renewcommand{\Im}{\mathop{\rm Im}\nolimits}

\newcommand{\beqa}{\begin{eqnarray}}
\newcommand{\eeqa}{\end{eqnarray}}

\title{
   Effective Action Approach for Preheating
}

\author[a,b]{Bin Xu} 
\author[a]{and Wei Xue}
\affiliation[a]{Department of Physics, University of Florida,
  Gainesville, FL 32611, USA}
\affiliation[b]{Center for High Energy Physics, Peking University, Beijing 100871, China}

\emailAdd{binxu@pku.edu.cn}
\emailAdd{weixue@ufl.edu}

\abstract{
We present a semiclassical non-perturbative approach for calculating the preheating process at the end of inflation. 
Our method involves integrating out the decayed particles within the path integral framework and subsequently determining 
world-line instanton solutions in the effective action. This enables us to obtain the effective action of the inflaton, with its imaginary part 
linked to the phenomenon of particle creation driven by coherent inflaton field oscillations. 
Additionally, we utilize the Bogoliubov transformation to investigate the evolution of particle density within the medium 
after multiple inflaton oscillations. 
We apply our approach to various final state particles, including scalar fields, tachyonic fields, and gauge fields.
The non-perturbative approach provides analytical results for preheating that are in accord with previous methods. 
}

\begin{document}
\maketitle
\flushbottom

\section{Introduction}
\label{sec:intro}

Compelling evidence from cosmological observations points out the existence of an inflationary phase 
\cite{Guth:1980zm,Linde:1981mu,Albrecht:1982wi,Starobinsky:1980te,Mukhanov:1981xt}, a critical element in the history of the Universe.
Inflation refers to a primordial epoch at the beginning of the universe, characterized by a slowly varying vacuum energy.
This energy is typically attributed to a scalar field known as the inflaton, which accelerates the expansion of the Universe,
leading to a supercooled phase by the exponential growth.

In parallel, we have confidence that 
the Universe underwent a thermal history from electron-positron annihilation with temperature about ${\rm MeV}$.
To link the supercooled inflation phase with the thermal history, 
a crucial intermediary step named 
reheating \cite{Abbott:1982hn,Dolgov:1982th,Albrecht:1982mp,Traschen:1990sw,Dolgov:1989us,Kofman:1994rk,Shtanov:1994ce,Kofman:1997yn} 
becomes apparent. Reheating entails introducing interactions between the inflaton field and other fields, including standard model particles. 
We must have these couplings to terminate inflation by facilitating the decay of the inflaton field, thereby generating entropy in the Universe.
The decayed particles would reach thermal equilibrium through mutual interactions,
initiating the thermal history of the Universe.

However, the conventional notion of perturbative decay does not offer a comprehensive description of the reheating process.
This is particularly true when dealing with the decay of a coherent oscillating field at the end of inflation, which differs 
fundamentally from the decay of individual particles.
One intriguing possibility arises when the decayed particles are bosons, leading to Bose-enhanced production rates. 
More notably, 
a sizeable coupling between the inflaton and the decayed particles can give rise to non-perturbative particle production via
parametric resonances.
These enhancement effects typically occur before the perturbative decay, known as preheating. 
The concept of preheating was initially proposed in \cite{Traschen:1990sw}, and further detailed studies are given in
\cite{Kofman:1994rk,Shtanov:1994ce,Kofman:1997yn}~(see \cite{Allahverdi:2010xz} for a review).

Future experiments are opening new windows to explore the critical and phenomenologically rich reheating period, 
though observing reheating or preheating directly is challenging. The challenge arises from the fact that the process of thermalization 
tends to erase most of the memory of reheating.
Nonetheless, the signals that remain unthermalized with the plasma offer hope.
The reheating can yield discernible signatures in the form of 
gravitational waves \cite{Khlebnikov:1997di,Easther:2006vd,Easther:2006gt,Garcia-Bellido:2007fiu,Dufaux:2007pt,Dufaux:2008dn,
Dufaux:2010cf,Bethke:2013vca,Adshead:2018doq,Kitajima:2018zco,Bartolo:2016ami,Figueroa:2017vfa,Caprini:2018mtu,Bartolo:2018qqn,
Lozanov:2019ylm,Adshead:2019igv,Adshead:2019lbr}, 
non-Gaussianity \cite{Enqvist:2004ey,Enqvist:2005qu,Jokinen:2005by,Barnaby:2006cq,Barnaby:2006km,Kohri:2009ac,Chambers:2007se,
Chambers:2008gu,Bond:2009xx,Leung:2012ve,Leung:2013rza,Imrith:2019njf,Fan:2020xgh}, 
magnetic field \cite{Calzetta:2001cf,Davis:2000zp,Boyanovsky:2002wa,Boyanovsky:2002kq,Mazumdar:2008up,Diaz-Gil:2007fch,Diaz-Gil:2008raf}, 
primordial black holes \cite{Green:2000he,Bassett:2000ha,Suyama:2004mz,Martin:2019nuw}, 
topological defects \cite{Khlebnikov:1998sz,Tkachev:1998dc,Parry:1998de}, and more.

In this paper, we concentrate on the theoretical aspects of preheating
and present an alternative perspective on preheating
by employing an effective action method. 
Preheating typically involves the generation of particles when the oscillating inflaton field crosses the minimum of its potential. 
During this crossing, the frequencies of the newly created particles change non-adiabatically, leading to resonant particle production.
The production can be comprehended through the examination of the mode functions of the decayed particles, which can be solved either numerically or 
by employing the WKB method (as discussed in \cite{Kofman:1997yn}).
Analysis of the mode functions reveals that resonant production occurs within a narrow or broad range of momentum space
contingent upon the coupling strength between
the inflaton field and the daughter particle.
A small coupling results in a narrow instability band where the momentum of decayed particles is close to the mass of the inflaton, assuming the
bare mass of the decayed particles is significantly smaller. 
Resonance becomes much more efficient when the coupling is substantial, introducing significant quantum corrections to the Lagrangian.
This is a non-perturbative process involving multi-particle scatterings. Consequently, 
it becomes imperative to sum all relevant processes within the coherent inflaton background.  
An essential tool for this purpose is the quantum effective action of the inflaton field, derived by integrating out the decayed     
fields, which incorporates these non-perturbative quantum corrections.
This, in turn, allows us to deduce the resonantly produced number density of the decayed particles, 
serving as the motivation to explore
the one-loop effective action method to gain insight into the non-perturbative preheating process.

Additionally, our application of the effective action method to inflaton decay is inspired by a related phenomenon, the Schwinger mechanism 
\cite{Schwinger:1951nm}, which elucidates 
the non-perturbative pair production of electrons and positrons in a strong electric field.
Both preheating and Schwinger pair production are related to particle creation in some backgrounds:
one occurs in the presence of an electric field, and the other in an oscillating inflaton background.
In the context of the Schwinger mechanism, one can use the WKB method to solve the mode functions of decayed particles
\cite{Brezin:1970xf} or
employ the effective action method to derive the vacuum persistence rate \cite{Affleck:1981bma,Kim:2000un,Gies:2005bz,Dittrich:2000zu,Monin:2010qj}. 
The two methods complement each other. 
In the case of preheating, the literature already contains references to the WKB method (see e.g. \cite{Kofman:1997yn,Mukhanov:2005sc}). 
In this work, we embark on the exploration of the effective action method 
for understanding non-perturbative inflaton decays. 
This method has the potential to apply to various particle production processes in cosmology.

The structure of this paper is organized as follows. \Cref{sec:pert} describes inflaton perturbative decay through the effective action
method, serving as a preliminary exercise.
In \cref{sec:preheating}, 
we calculate the effective action by utilizing instanton solutions to illustrate the
non-perturbative particle creation during the preheating period and 
introduce the Bogoliubov transformation to account for the evolution of particle number density.
In \cref{sec:othermodel}, we apply our approach to other models, including tachyonic resonance and
gauge preheating. We conclude with a summary and discussion of our findings in \cref{sec:conclusion}.
Appendix includes the numerical and analytic methods for calculating the effective action 
in the presence of a periodic potential.

\section{Reheating -- perturbative inflaton decay}
\label{sec:pert}

In this section, we delve into the analysis of reheating, specifically focusing on perturbative inflaton decay.
We employ the effective action method as a preliminary example, which establishes the equivalence between 
the perturbative decay rate of a coherent oscillating inflaton and that of a particle with the same mass and coupling.

Let us consider a reheating model where an inflaton $\phi$ is a real scalar field with mass $M_\phi$.
It interacts with another real scalar 
field $\chi$ with mass $m_\chi$, and the interaction is characterized by a trilinear coupling $g$.
We can decompose the action into three parts: 
the free action of the inflaton, $S_0 [ \phi ]$, the free action of the $\chi$ field, $S_0 [ \phi ]$, and the
interaction part, $S_I[ \phi, \chi]$. The action takes the following form 
\begin{equation}
	S[ \phi, \chi] = S_0 [ \phi ]  + S_0[  \chi]  + S_I[ \phi, \chi] = \int d^4 x\, \left[ -\frac{1}{2}\phi(\Box+M_\phi^2)\phi-\frac{1}{2}\chi(\Box+m_\chi^2)\chi-\frac{g}{2}\phi\chi^2 \right] \ ,
\end{equation}
where the D'Alembert operator $ \Box = \partial_\mu \partial^\mu$. 
The rate at which an inflaton particle decays at rest can be deduced using a standard S-matrix approach, 
\begin{equation}
	\Gamma(\phi\to\chi\chi)
	=\frac{g^2}{32\pi M_\phi}\sqrt{1-4\frac{m_\chi^2}{M_\phi^2}}\, \theta(M_\phi-2m_\chi)   \ .
\label{eq:phiparticledecay}
\end{equation}

In the case of reheating, when we neglect the expansion of the universe and the backreaction of inflaton
decays, the classical solution of a homogeneous inflaton field with the maximum amplitude $\phi_0$
is given by the field equation derived from the free inflaton action $S_0 [ \phi ]$,
\begin{equation} 
   \phi (t) = \phi_0  \, \sin ( M_\phi t ) \ .  
\label{eq:phit}
\end{equation} 
Here, the inflaton is coherently oscillating, and the effective action approach is a suitable tool to 
consider the decay of a background field. We 
expand the interaction action to the second power, and then integrate the $\chi$ fields using the path integration, 
leading to the effective action of $\phi$, 
\begin{equation}
   \begin{split}
	\int\mathcal{D}\phi e^{iS_{\text{eff}}[\phi]}&=\int\mathcal{D}\phi\mathcal{D}\chi e^{iS_0[\phi]+iS_0[\chi]+iS_I[\phi,\chi]}\\
	&=\int\mathcal{D}\phi e^{iS_0[\phi]}(1+i \langle S_I \rangle-\frac{\langle S_I^2 \rangle}{2}+\cdots) \, , 
   \end{split}
\end{equation}
where $\langle S_I \rangle  = \int\mathcal{D}\chi e^{iS_0[\chi]} S_I$ and $\langle S_I^2 \rangle$ is similar.
We then treat the inflaton field $\phi$ as a classical background to eliminate $\int \mathcal{D}\phi$ and obtain the form of the effective action,
\begin{equation}
\begin{split}
	S_{\text{eff}} [\phi]
		=&-\int d^4x\frac{1}{2}\phi(\Box+M^2)\phi-\int d^4x\frac{g}{2}\phi(x)\Delta_F(x,x)
      \\
      &+i\int d^4x_1d^4x_2\left(\frac{g}{2}\right)^2\phi(x_1)\phi(x_2)\Delta_F^2(x_1,x_2)
   \label{eq:Seff}
\end{split}
\end{equation}
where $\Delta_F(x_1,x_2)$ is the Feynman propagator,  
\begin{equation}
	\Delta_F(x_1,x_2)=\int\frac{d^4k}{(2\pi)^4}\frac{i}{k^2-m^2+i\epsilon}e^{ik(x_1-x_2)} \ . 
\end{equation}
The vacuum persistence probability is calculated using the S-matrix amplitude with the effective action 
$ |\langle \phi | e^{i S_{eff}} | \phi  \rangle  |^2 = e^{ - 2 \Im S_{eff} ( \phi) }$, which is related to the coherent inflaton decay rate.
The leading imaginary part of the effective action is given by second order in $g$, and we use the background $\phi(x)$ given in \cref{eq:phit}
to write the imaginary part of the effective action,
\begin{equation}
\begin{split}
	\Im S_{\text{eff}}
	&=\frac{g^2}{4}\int d^4x_1d^4x_2 \, \phi(x_1)\phi(x_2) \, 
      \Im \left[i \Delta_F^2(x_1,x_2)\right] \\
	&=\frac{g^2 \, \phi_0^2 \, VT}{128\pi}\sqrt{1-\frac{4m_\chi^2}{M_\phi^2}} \, \theta(M_\phi-2m_\chi) \ .
\end{split}
\end{equation}
Here, we use the imaginary part
of $ i \Delta_F^2(x_1,x_2) $, 
\begin{equation}
    \Im[i \Delta_F^2(x_1,x_2)] =\frac{1}{16\pi}\int\frac{d^4k}{(2\pi)^4}e^{ik(x_1-x_2)}\sqrt{1-\frac{4 m_\chi^2}{k^2}} \, \theta(k^2-4m_\chi^2)  \ .
\end{equation}
\footnote{By dimensional regularization, one can show that
\begin{equation}
	\Delta_F(x,x)=-\frac{m_\chi^2}{8\pi^2}\left[\frac{1}{\epsilon}+\text{constant}-\frac{1}{2}\ln\left(\frac{m_\chi^2}{4\pi\mu^2}\right)\right]
\end{equation}
\begin{equation}
	i\Delta_F^2(x_1,x_2)=\int\frac{d^4k}{(2\pi)^4}e^{ik(x_1-x_2)}\frac{1}{8\pi^2}\left[\frac{1}{\epsilon}+\text{constant}
      -\frac{1}{2}\int_{0}^{1}dx\ln\left(\frac{m_\chi^2-k^2x(1-x)-i\epsilon}{4\pi\mu^2}\right)\right]
\end{equation}
where $\mu$ is the renormalization scale.}
The vacuum persistence probability leads to the inflaton background's total decay rate per unit time per unit volume, 
\begin{equation}
	R(\phi\to\chi\chi)=\frac{2\Im S_{\text{eff}}}{VT }
      = n_\phi \frac{g^2}{32\pi M_\phi}\sqrt{1-\frac{4m_\chi^2}{M_\phi^2}}\theta(M_\phi^2-4m_\chi^2) \ . 
\end{equation}
The number density of the inflaton field is given by 
\begin{equation}
	n_\phi=
      \frac{1}{2M_\phi}(\dot{\phi}^2+M_\phi^2\phi^2)\simeq\frac{1}{2}M_\phi \phi_0^2 \ . 
\end{equation}
This result is consistent with the 
number density times the particle decay rate, i.e., $R(\phi\to\chi\chi) = n_\phi\Gamma( \phi \to \chi \chi )$.
Hence, the inflaton decay rate deduced from the effective action approach is equivalent to the particle decay rate 
given in \cref{eq:phiparticledecay}.

\section{Parametric Resonance }
\label{sec:preheating}

In the period of the end of inflation, non-perturbative effects come into play when the inflaton condensation oscillates near the minimum of its potential. Two well-established mechanisms are of particular interest: parametric resonance
and tachyonic instability. In this section, we will delve into the phenomenon of parametric resonance 
using the effective action approach and then explore the tachyonic instability in the following section.

We begin by considering an inflaton model in which the inflaton field couples with a scalar field $\chi$ via a quartic interaction term,
represented as $\frac{g^2}{2} \phi^2 \chi^2$. The action for this model takes the following form, 
\begin{equation}
	S[ \phi, \chi] = \int d^4 x\, \left[ -\frac{1}{2}\phi(\Box+M_\phi^2)\phi
         -\frac{1}{2}\chi(\Box+m_\chi^2)\chi-\frac{g^2}{2}\phi^2\chi^2 \right] \, .
\end{equation}
Given the oscillation behavior of the inflaton condensation as described in \cref{eq:phit}, 
the field equation for $\chi$ can be mapped to the Mathieu equation, which manifests 
certain instability regions in momentum space ($k$-space). Depending on the strength of the coupling, 
when the coupling is relatively small ($g  \ll  m / \phi_0 $), 
resonance occurs in a narrow instability band. Conversely, in the case of large coupling ($g   >  m / \phi_0 $), 
the resonance effect spans a broader band, leading to more efficient inflaton decays. 
The broad band resonance is non-perturbative.
It can be understood that the rate of the process involving the annihilation of multiple inflatons into $\chi$ particles 
($n \phi \to 2 \chi$) is comparable to or even exceeds the rate of the process $2\phi \to 2\chi$.
Consequently, in this regime, the decay is inherently non-perturbative, resulting in the opening of broad resonance bands.
Note that our primary focus here is on the effects of broad resonance.
We have checked that the narrow resonance using the effective action method gives an inflaton background decay rate with the Bose enhancement, 
consistent with the narrow resonance study in \cite{Kofman:1997yn,Mukhanov:2005sc}. 

Since the non-perturbative inflaton decay yields two $\chi$ particles as final states, 
we evaluate this process using 
the one-loop effective action approach by integrating out the $\chi$ fields, 
\begin{equation}
	\int\mathcal{D}\phi e^{iS_{\text{eff}}[\phi]} =
      \int\mathcal{D}\phi\mathcal{D}\chi e^{iS_0[\phi]+iS_0[\chi]+iS_I[\phi,\chi]} \, .
\end{equation}
Subsequently, with the effective action of $\phi$ in hand, we employ the Bogoliubov
transformation to investigate the stimulated particle production and gain insights into the evolution of the number density of $\chi$.

\subsection{One-loop effective action}

The oscillating background field $\phi(t) = \phi_0(t) \sin (M_\phi t ) $ imparts a time-dependent frequency $\omega(t)$ to the $\chi$ particle. 
The adiabatic condition $\dot\omega(t) / \omega^2(t)$ is significantly violated when the oscillating inflaton passes the minimum of the 
potential. At the minimum, the field value vanishes at $t_j = j \, \pi  / M_\phi$, where $j$ is an integer.
However, for majority of time, $\dot\omega(t) / \omega^2(t) \ll 1 $, ensuring the conservation of the number of $\chi$ particles
when inflaton is not 
near the potential minimum. Therefore, it is 
sufficient to investigate particle creation when $\phi(t)$ is close to $0$, and $\phi^2(t)$ takes an approximate form, 
\begin{equation}
   \phi^2(t) \simeq \phi_0^2 (t_j ) \, M_\phi^2  ( t - t_j)^2 , \quad \quad \quad  t_j  = \frac{ j \, \pi}  { M_\phi } \, .
\label{phi2app}
\end{equation}
This approximation simplifies the calculation of the effective action method and 
yields analytic results. 
Additionally, in a flat space, $\phi_0$ is a constant, while in an expanding universe, the time-dependence of $\phi_0(t)$ 
needs to be taken into account. \Cref{phi2app} remains a valid approximation as long as the variation of $\phi_0(t)$ 
is small compared to the inflaton mass $M_\phi$.
Since the time duration of preheating is small compared to the Hubble time scale,
these assumptions are valid and self-consistent during the preheating process \cite{Kofman:1997yn,Mukhanov:2005sc}.

There is one more complication in the $\chi$ particle production rate. The rate is momentum-dependent, 
and due to medium effects, the rates for different modes vary with time according to their occupation number. 
When integrating out all $\chi$ particle modes in the path integration, we will attain the vacuum persistence 
probability, which is not a useful quantity to understand the momentum-dependent preheating process. 
Instead, we factorize the $\chi$ field into $k$ modes
\begin{equation}
    \chi(t,x)=\int\frac{d^3k}{(2\pi)^3}\chi_\mathbf{k}(t) \, e^{i\mathbf{k}\cdot \mathbf{x}} \, ,
\end{equation}
and derive the decay probability for individual $k$ mode via the effective action method.

Since the interaction terms are quadratic in the field $\chi$, different $k$ modes are not coupled together. 
This is evident from the action,
\begin{equation}
    S_0[\chi]+S_I[\chi,\phi] =
         \int dt 
      \frac{d^3k}{(2\pi)^3}
      (-\frac{1}{2})\chi^*_\mathbf{k} \left( 
         \partial_t^2+\mathbf{k}^2+m^2_\chi +g^2\phi^2(t) \right)
         \chi_\mathbf{k}
      \, .
\end{equation}
For convenience, we discretize the momentum ${\bf k}$, and then integrate out the $\chi$ field mode by mode in the Feynman path integration, 
\begin{equation}
    e^{i S_{\rm eff}[\phi]} = \int \mathcal{D}\chi e^{iS_0[\chi]+iS_I[\chi,\phi]}
      =  {\cal  N } \prod_\mathbf{k} \frac{1}{\sqrt {\det(\partial_t^2+\mathbf{k}^2+m^2_\chi+g^2\phi^2(t) )} }\, , 
\label{eq:Seff}
\end{equation}
where ${\cal N}$ is a $\mathbf{k}$- and $t$-independent normalization constant. 
The path integral gives the effective action of the inflaton field 
with the summation over all $\bf k$, 
\begin{equation}
    S_{\rm eff}= \sum_\mathbf{k} S_\mathbf{k}
\end{equation}
We compute the effective action for a given momentum ${\bf k}$ by introducing the Schwinger proper time $s$, 
\begin{equation}
   S_\mathbf{k} = 
     \frac{i}{2}  \int {\rm d}t  \, \langle t|\ln(\partial_t^2+\mathbf{k}^2+m^2_\chi+g^2\phi^2)|t\rangle 
      =- \frac{ i}{2}  \int_0^\infty \frac{ds}{s} e^{-is(\mathbf{k}^2+m^2_\chi)}
      \int {\rm d}t   \langle t|e^{-i\hat{H}s}|t\rangle
   \label{eq:Sk}
\end{equation}
where $\hat{H}=\partial_t^2+g^2 \phi^2(t)$. Here, we sum over a one-particle Hilbert space spanned by $| t \rangle$ via the integration 
$\int {\rm d} t$. 
To obtain the effective action, we first compute the amplitude of 
the particle propagating to the same state after the proper time $s$.
Several approaches can be used to calculate the time $t$ integration in \cref{eq:Sk}. 
The world-line instanton approach is presented here. This approach is used in this 
section on parametric resonance and the next section on tachyonic instability.
An alternative method of summing over all the eigenstates is presented after the instanton approach.

\subsubsection*{Instanton approach}

The instanton approach leverages the Schwinger proper time and finds the 
semi-classical instanton solutions to the amplitude for quantum mechanical states evolving in the Schwinger time.
This approach has been successfully applied to the problems involving electron-positron pair production in 
the presence of electromagnetic backgrounds \cite{Affleck:1981bma,Kim:2000un,Gies:2005bz,Dittrich:2000zu,Monin:2010qj}.

Starting from the effective action formula given in \cref{eq:Sk}, we rotate both time $t$ and Schwinger time $s$ into 
Euclidean ones, $t \to i \tau$ and $ s \to - i s$. 
\footnote{This rotation into Euclidean Schwinger time alone would also yield the same result.}
The expectation value of the evolution operator is then evaluated by summing over the histories of $\tau(s)$, 
\begin{equation}
    \langle \tau |e^{-\hat{H}s}| \tau \rangle=\int_{\tau(0)= \tau(s)}
   \mathcal{D}x e^{-S} 
      \simeq e^{-S_0}\det\left(-\frac{1}{2}\frac{d^2}{ds^2}+V''(\tau_{cl} ) \right)^{-\frac{1}{2}} \, .
   \label{eq:evolutionO}
\end{equation}
Here $\tau_{cl}$ represents a stationary point of $S$, signifying a classical trajectory of instanton. The prime symbol indicates
differentiation with respect to $\tau$,
and $S_0$ denotes the classical action of the instanton,
\begin{equation}
    S_0=\int_0^s ds' L=\int_0^s ds' (\frac{1}{4} ( \frac{ d \tau }{ d s'} )^2+V (\tau ) ) \, .
\label{eq:S0cl}
\end{equation}
For the approximation of the inflaton field $\phi(t)$ oscillating near the minimum, we have a quadratic potential,
\begin{equation}
    V(\tau)
   =-g^2 M_\phi^2 \phi_0^2\tau^2 =  - \omega^2 \tau^2 \, ,    
\end{equation}
where we define 
\begin{equation}
   \omega \equiv  g  M_\phi \phi_0 \, .
\end{equation}
We solve the equation of motion for $\tau(s')$ with the boundary condition 
$\tau (0) = \tau ( s ) = \tau_0$, and find the classical orbit, 
\begin{equation}
    \tau_{cl} (s')=\frac{\tau_0}{\cos\omega s}\cos(2\omega s'-\omega s) \, .
\end{equation}
Hence, when $\tau_{cl} (s')$ is inserted into \cref{eq:S0cl}, the classical action takes the following form,
\begin{equation}
    S_0=-\omega\tau_0^2\tan\omega s \, .
   \label{eq:S0value}
\end{equation} 
The determinant in \cref{eq:evolutionO} can be evaluated using a classical action $S_{ab}$ 
with the boundary $\tau(0) = \tau_a$ and $\tau(s) = \tau_b$,
\begin{equation}
    \det\left(-\frac{1}{2}\frac{d^2}{ds^2}-2\omega^2\right) = -\frac{\mathcal{N}}{2}\left(\frac{\partial^2S_{ab}}{\partial\tau_a\partial\tau_b}\right)^{-1} = \mathcal{N} \frac{\sin(2\omega s)}{2\omega}
   \label{eq:det}
\end{equation}
Alternatively, the determinant can be evaluated using the Gelfand-Yaglom formula \cite{Gelfand:1959nq}. 
The normalization factor $\cal N$ is attained for the free case 
$V(\tau) = 0$, which gives ${\cal N}  = 4 \pi $. According to \cref{eq:evolutionO,eq:S0value,eq:det},
the imaginary part of the effective action of \cref{eq:Sk} is therefore
\begin{equation}
   \begin{split}
     {\rm Im} S_{\bf k}  &= 
       -   {\rm Re}  \, \frac{1}{2}  \int d\tau  
         \int_0^\infty \frac{ds}{s} e^{-s(\mathbf{k}^2+m^2_\chi)}
      e^{\omega\tau^2\tan\omega s} 
      \sqrt{\frac{2\omega} 
      {4\pi \sin 2\omega s}}
      \\
   &= 
      -\, {\rm Re}  \, \frac{1}{2} \int_0^\infty \frac{ds}{s} e^{-s(\mathbf{k}^2+m^2_\chi)} \left(\frac{i}{2\sin\omega s}\right)\\
   \end{split}
\label{eq:Seff_instanton}
\end{equation}
In the second line, we integrate the imaginary time $\tau$.
We evaluate the integration of $s$ by applying the residual theorem, but the integration is divergent at $s=0$, and the pole at this point 
is not a simple pole. We regulate the divergence by subtracting the pole at $s= 0$. After regularization, we integrate the Schwinger time
$s$, and find that, 
\begin{equation}
   \begin{split}
     {\rm Im} S_{\bf k}  &= 
      -\, {\rm Re}  \, \frac{1}{2} \int_0^\infty \frac{ds}{s} e^{-s(\mathbf{k}^2+m^2_\chi)} 
      \left(\frac{i}{2\sin\omega s} - 
      \frac{i}{2 \omega s} \right)  \\
    &= 
      \frac{1}{4} \sum_n \frac{(-1)^{n-1}}{n} {\rm e}^{-\frac{n\pi}{ \omega }(\mathbf{k}^2+m^2_\chi)} 
      \\
         &= \frac{1}{4} \ln\left[1+ \exp ({-\frac{\pi(\mathbf{k}^2+m^2_\chi)}{g M_\phi \phi_0}}) \right]
      \label{eq:ImS_reg}
   \end{split}
\end{equation}
where we sum over all positive integers of $n$.
Therefore, during half an oscillation period, vacuum persistence probability is
\begin{equation}
   P_{vac}
      = {\rm e}^{ - \sum_{\bf k } 2 {\rm Im} S_{\bf k}}    =
 \prod_{\bf k} \frac{1} {\sqrt{1+ \exp ({-\frac{\pi(\mathbf{k}^2+m^2_\chi)}{g M_\phi \phi_0}})}}  
  = \prod_{ \{ {\bf k}, - {\bf k }\} } \frac{1} {1+ \exp ({-\frac{\pi(\mathbf{k}^2+m^2_\chi)}{g M_\phi \phi_0}})}
   \, .
\label{eq:Rvac}
\end{equation}
This quantity is the amplitude square, directly corresponding to a probability. 
Also, ${\bf k}$ represents the momentum of $\chi$ particle. It's important to note that ${\bf k}$ and $-{\bf k}$ correspond to 
the same final states due to pair particle production and momentum conservation.
In the final equation above, we combine the contributions of ${\bf k}$ and $-{\bf k}$ to account for 
the pair production of these modes.
This equation reveals the probability of creating $k$-mode after the first oscillation,
\begin{equation}
   P_k( \phi \to \chi \chi ) 
          = {\rm e}^{{-\frac{\pi(\mathbf{k}^2+m^2_\chi)}{g M_\phi \phi_0}}} \, ,
\label{eq:Rk}
\end{equation}
Note that $P_{vac} + \sum_{\bf k} P_k \simeq 1$ if $P_{k}$ are small quantities. 
But the probability of creating $k$-mode $P_k$ in the above equation is more accurate, and it does not require that $P_k$ is small, 
which will become clear from the Bogoliubov transformation in \cref{sec:bogoliubov}.

A natural question arises regarding the applicability of the instanton approach without making the near potential minimum approximation.
In the appendix, the instanton method is applied to a periodic potential without this approximation, and 
the effective action is numerically calculated by truncating the Fock space.

\subsubsection*{Eigenstates}

With the approximation of the inflaton field $\phi(t)$ as described in \cref{phi2app}, the Hamiltonian $\hat{H}$ takes on a form analogous to 
harmonic oscillators with an imaginary frequency.
The time integration can be recast as a summation of all the energy eigenstates, giving 
\begin{equation}
    \int {\rm d}t \langle t| {\rm e}^{-i \hat{H} s }| t \rangle
         =\sum_n e^{-is E_n}  =  
         \sum_n {\rm e}^{-2\omega s(n+\frac{1}{2})}
         =  \frac{1}{2\sinh(\omega s)} \, .
\end{equation}
We regulate the divergence by straightforwardly excluding the pole at $s=0$, as shown \cref{eq:ImS_reg}.
Then, the imaginary part of the effective action is expressed as 
\begin{equation}
      {\rm Im} S_{\bf k} = 
      -  \, {\rm Re} \, \frac{1}{2}\int_0^\infty \frac{ds}{s} 
         {\rm e}^{-is(\mathbf{k}^2+m^2_\chi)} \left(
         \frac{1}{2\sinh\omega s}
         -\frac{1}{2\omega s}\right) \, .
\end{equation}
We calculate this integration by rotating $s \to - i s$ and applying the residual theorem, resulting in the same outcome 
as shown in \cref{eq:ImS_reg}. Consequently, we obtain the same probability for $\chi$ particle creation.

\subsection{Bogoliubov transformation}
\label{sec:bogoliubov}

The probabilities for particle creation in vacuum (\cref{eq:Rk}) and the vacuum persistence (\cref{eq:Rvac})
are only applicable during the initial oscillation of the inflaton. After a few oscillations, the number of $\chi$ particles grows exponentially 
due to the Bose enhancement. Thus, we must 
consider the medium effects on the inflaton's parametric decay. 
Still, we assume that the parametric resonance timescale 
is much shorter than the Hubble time, 
and the $\chi$ particles have neither thermalized nor decayed.
Under these conditions, we can elucidate the growth in the number of $\chi$ particles 
through the Bogoliubov transformation. 
Interestingly, this growth rate
relies on the parametric vacuum decay rate, as previously calculated, and the number of inflaton oscillations.

With the aid of the Bogoliubov transformation, we map the annihilation and creation operators in one oscillation 
to the ones in the previous oscillation, 
\begin{equation}
    a_\mathbf{k}(t+T)
      = \mu^*_{k} a_\mathbf{k}(t)+\nu_{k} a_{-\mathbf{k}}^\dagger(t)
\end{equation}
where $T =  \pi / M_\phi$ represents half of an inflaton oscillation period. 
The Bogoliubov coefficients, $\mu_{k}$ and $\nu_k$, 
satisfy the relation, 
\begin{equation}
    |\mu_{k}|^2-|\nu_{k}|^2=1 \, .
\end{equation}
Considering that the $\chi$ particles are pair-produced, and scatterings among $\chi$ particles are at a later stage, we decompose
the vacuum state at time $t$ as a superposition of states at $t+ T$, taking into account the produced pairs,   
\begin{equation}
    |0;t\rangle=
      \prod_{  \{ \mathbf{k}, - {\bf k} \}  }
      \frac{1}{\mu_k}
         \exp\left[\frac{\nu_k}   
         {\mu_k}a^\dagger_\mathbf{k}(t+T)a^\dagger_{-\mathbf{k}}(t+T)\right]|0;t+T\rangle \, .
\end{equation}
This decomposition aligns with the definition of the vacuum state at time $t$, where $a_\mathbf{k}(t)|0;t\rangle=0$,
and the prefactor derives from its normalization.
Let's consider the initial oscillation, 
during which the $\chi$ state remains in the vacuum. The vacuum persistence probability, as determined by the 
Bogoliubov transformation, should coincide with \cref{eq:Rvac},
\begin{equation}
    |\langle 0;T|0;t = 0 \rangle|^2= 
      \prod_{   \{ \mathbf{k}, - {\bf k} \}  }\frac{1}{|\mu_k|^2} 
       = \prod_{   \{ \mathbf{k}, - {\bf k} \}  }\frac{1}{ 1+ |\nu_k |^2} 
         =\prod_{   \{ \mathbf{k}, - {\bf k} \}  }\frac{1}{1+e^{-\frac{\pi(\mathbf{k}^2+m^2_\chi)}{g\Phi M}}} \,  .
\end{equation}
This yields the absolute value of $\nu_k$,
\begin{equation}
  |\nu_{k}|^2=e^{-\frac{\pi(\mathbf{k}^2+m^2_\chi)}{g\Phi M_\Phi}}  \, ,
\end{equation}
which justifies that $P_k$ in \cref{eq:Rk} is an accurate result.
To assess stimulated particle pair production in a medium with a non-zero average occupation number, we introduce the density matrix $\rho$,
\begin{align}
    n_k(t+T)&=Tr[a^\dagger(t+T)a(t+T)\rho]
      \nonumber
      \\
    &=|\mu_{k}|^2 {\rm Tr} [a^\dagger(t)a(t)\rho]+|\nu_{k}|^2 {\rm Tr} [a(t)a^\dagger(t)\rho]+2 {\rm Re}\nu_{k}^*\mu_{k} {\rm Tr} [a^2(t)\rho]
      \nonumber
      \\
    &\approx (1+2|\nu_{k}|^2+2|\mu_{k}||\nu_{k}|\cos\theta_{k}) n_k(t)
\end{align}
where $\theta_{k}$ represents an arbitrary phase. We consider $n_{k}$ to be large so that all the traces of the density matrix
in the above equation gives an approximation of the same result, $n_k$.
When $n_{k}$ is substantial, the occupation number undergoes exponential growth,
\begin{equation}
    n_{k}(NT)\sim e^{N\ln(1+2|\nu_{k}|^2+2|\mu_{k}||\nu_{k}|\cos\theta_{k})}
\end{equation}
This result aligns with previous findings, which were obtained using the WKB method \cite{Kofman:1997yn,Mukhanov:2005sc}.
The outcome is a function of $\nu_k$, as governed by the vacuum persistence probability via the effective action method. 
The phase $\theta_k$, which can be evaluated using the WKB method, becomes larger than $\pi$ and rather complicated time-dependent 
in an expanding universe. 
Consequently, it can be approximated as 
a random phase, introducing stochasticity to the $\chi$ particle number growth. 
This allows us to average the random phase,
which yields the evolution of number density. The final result 
relies on both the vacuum persistent probability and the number of inflaton oscillations.

\section{Tachyonic Instability}
\label{sec:othermodel}

In this section, we will apply the effective action method to study tachyonic instability in other preheating models: 
one is tachyonic preheating 
\cite{Greene:1997ge,Felder:2000hj,Felder:2001kt,Shuhmaher:2005mf,Dufaux:2006ee,Abolhasani:2009nb} and 
the other is gauge boson preheating via a Chern-Simons coupling \cite{Deskins:2013dwa,Garcia-Bellido:2003wva,Adshead:2015pva,McDonough:2016xvu,
Adshead:2016iae,Adshead:2018doq}. 
The effective action method provides analytic results
consistent with the previous findings.

\subsection{Tachyonic preheating}

Following \cite{Dufaux:2006ee},
we introduce a bosonic trilinear coupling between the inflaton $\phi$ and its daughter particle $\chi$ as a realization of
tachyonic preheating, 
\begin{equation}
    \mathcal{S [\chi]}= \int d^4x \left[-\frac{1}{2}\chi(\Box+m_\chi^2)\chi  -  \frac{g}{2}\phi\chi^2 \right] \, .
\end{equation}
Similarly, we assume that the preheating timescale is significantly shorter than the Hubble time, enabling us to disregard the expansion 
of the universe. Then the field equation for $\chi_{\bf k}$ takes the form
\begin{equation} 
    \ddot\chi_{\bf k}+ \omega_k^2 \chi_{\bf k}=0  \, ,  
   \quad \omega_k^2  = ( k^2+m_\chi^2+g\phi_0(t)  ) \, .
\end{equation}
With the inflaton background evolving as $\phi_0(t) = \phi_0 \sin ( M_\phi t ) $, the field equation above can be reduced to 
the canonical Mathieu equation. When the coupling is large, with $g\phi_0  > M_\phi^2$, we enter the broad resonance region. 
However, modes with negative $\omega_k^2$ undergo tachyonic instability, 
which cannot be approximated as waves scattering from a parabolic potential, as in the broad resonance case.
Interestingly, the effective action doesn't tell the difference between both cases and treats them as a whole.

Our strategy parallels the one employed in the previous section.
In the tachyonic resonance model, we first integrate out $\chi_{\bf k}$ modes using the path integral to determine the effective action of the inflaton,
as in \cref{eq:Seff,eq:Sk}.
Second, we evaluate the Schwinger time evolution by finding the world-line instanton solution shown in \cref{eq:evolutionO}.
One minor distinction from the parametric resonance calculation is that here, we do not perform a rotation of the time $t$ into the Euclidean time. 
However, we have verified that using both real-time and Euclidean time yields the same result.
In the tachyonic resonance model, the classical action of the instanton is given as 
\begin{equation}
    S_0= 
      \int_0^s ds' (\frac{1}{4} ( \frac{ d t }{ d s'} )^2  + V (t)) = 
      \int_0^s ds' (\frac{1}{4}(  \frac{ d t }{ d s'} )^2 -g\phi_0\sin(M_\phi t)) \, .
\end{equation}
One of the saddle point solutions occurs at the peak of the potential $V(t)$, with $\dot{t} = 0 $, 
representing the mode with the largest negative mass square. Here, we approximate the sine function near the potential's peak
as a quadratic term, 
\begin{equation}
    g\phi_0\sin (M_\phi t)\approx -g \phi_0+\frac{1}{2}g\phi_0  M_\phi^2( t  - t_j )^2 \, , \quad t_j = \frac{(2j+ \frac{1}{2} )\pi}{M_\phi} \,  ,
\end{equation}
where $j$ is an integer.   
The saddle-point solution yields the classical action,
\begin{equation}
   S_0 = -  \omega t^2\tan\omega s - g\phi_0 s  \, ,
\end{equation}
with $\omega=\sqrt{\frac{g\phi_0 M_\phi^2}{2}}$. 
Combing the classical action and the determinant, which aligns with  
with the one in broad resonance case in \cref{eq:det}, we obtain the Schwinger time evolution operator, 
\begin{equation}
    \langle t|e^{-\hat{H}s}|t\rangle 
   \simeq e^{-S_0}\det(-\frac{1}{2}\frac{d^2}{ds^2}+V'')^{-\frac{1}{2}}=
\sqrt{\frac{\omega}{2\pi\sin 2\omega s}}e^{\omega t^2\tan\omega s+g \phi_0 s} \ .
\end{equation}
By putting these solutions together and conducting a similar integration as in \cref{eq:Seff_instanton,eq:ImS_reg}, we determine that 
the imaginary part of the effective action takes the form 
\begin{equation}
     {\rm Im}  S_\mathbf{k}
      =
      \frac{1}{4}\ln\left[1+\exp ({-\frac{\pi(\mathbf{k}^2+m^2_\chi-g\phi_0)}{\omega }})\right] \, .
\end{equation}
This imaginary part of the effective action gives the probability for $\chi$ particle creation after the first oscillation,
\begin{equation}
    P_k  =  |\nu_k|^2 = \exp(-\frac{\pi (\mathbf{k}^2+m^2_\chi-g\phi_0)}
      {  \sqrt{\frac{g\phi_0 M_\phi^2}{2}}  })  \, .
   \label{eq:nk_tp}
\end{equation}
It's noteworthy that the prefactor in the exponent, equal to $\pi$, is in close agreement with the numerical value found in \cite{Dufaux:2006ee}, 
which is approximately $3.4$.
The time evolution of the number density $n_k$ 
can be obtained through the Bogoliubov transformation as presented in \cref{sec:bogoliubov}, with the phase $\theta_k$ remaining indeterminate 
in this context.

\subsection{Gauge preheating}

In this preheating model, we consider a massive or massless gauge boson, denoted as $A_\mu$, which couples to 
the inflaton via a Chern-Simons operator. This coupling arises due to an approximate shift symmetry of inflaton.
The Lagrangian for this model can be expressed as
\begin{equation}
    \mathcal{L}= -\frac{1}{4}F_{\mu\nu}F^{\mu\nu}
      +\frac{1}{2}m^2_A A_\mu A^\mu
      -\frac{1}{4 \Lambda}\phi F_{\mu\nu}\tilde F^{\mu\nu} \, .
\end{equation}
Here the Chern-Simons coupling involves the field strength $F_{\mu \nu} = \partial_\mu A_\nu - \partial_\nu A_\mu$ 
and its dual $\tilde F^{\mu\nu}  = \frac{1}{2} \epsilon^{\mu \nu \rho \sigma } F_{\rho \sigma}$, and the coupling
strength is represented by a dimensional quantity $\frac{1}{ \Lambda}$.
A massive gauge boson with a constraint $\partial_\mu A^\mu = 0$ can be decomposed into helicity basis components: 
two transverse modes, $h = \pm 1$, and one longitudinal mode, $h = 0$. Consequently, the gauge boson action takes the following form 
\begin{equation}
    S[A]=\sum_{\lambda=\pm,0} \int dt  \int\frac{d^3 \bf k}{(2\pi)^3}
       (-\frac{1}{2})A_{\bf k}^{\lambda*}(\partial_t^2+\mathbf{k}^2+m^2_A-\frac{k}{ \Lambda}\dot\phi)A_{\bf k}^{\lambda} \, .
\end{equation}
A significant particle production occurs around the minimal of $-\frac{k}{ \Lambda}\dot\phi$, indicating that 
the $A^+_{\bf k}$ modes predominantly produced when $\phi= 0 $ and $\dot{\phi} >0$. 
Similarly, the $A^-_k$ modes predominantly produced when $\phi= 0 $ and $\dot{\phi} <0$. 
As the longitudinal mode does not couple to the inflaton 
and the action is equivalent to a free theory, the longitudinal mode production is not expected during the process.

The world-line instanton solution in this model also possesses a saddle point at the peak of the potential,
\begin{equation}
V(t) =  -\frac{k}{ \Lambda}\dot\phi(t) =  -\frac{k}{ \Lambda}\phi_0 \cos(M_\phi t) \, .
\end{equation}
We still employ the quadratic potential approximation for this case. 
When the inflaton crosses the origin for the first time with $\dot{\phi} >0$, the probability that the $A^+_{\bf k}$ modes are produced is given as  
\begin{equation}
    P_k^+  = \exp(-\frac{ \pi ( \mathbf{k}^2+m^2_A-\frac{k \phi_0 M_\phi }{\Lambda})} { \sqrt{ \frac{k \phi_0 M_\phi^3 }{2 \Lambda}   }  }   )
      \, .
\end{equation}
The same result applies to $A^-_{\bf k}$ modes, but the production occurs when $\dot{\phi} <0$, crossing the minimum of the potential. 
The particle creation probability is found to be numerically consistent with previous literature \cite{Adshead:2015pva}.

\section{Conclusion}
\label{sec:conclusion}

In this paper, we have employed the effective action approach to investigate the non-perturbative inflaton decays in an oscillating 
inflaton background. By focusing on the imaginary part of the effective action, we obtain the probability of the inflaton 
decaying to its coupled particle. This approach allows for the examination of both perturbative decays, such as the reheating
process and narrow resonances in preheating, and non-perturbative decays. Here, we concentrate on non-perturbative particle production through
broad parametric resonance or tachyonic instability.

To arrive at our results, we have integrated out decayed fields in the Path Integral, employed the world-line instanton method
to identify the semiclassical solutions of the effective action and subsequently determine the particle decay probabilities in the vacuum.  
Moreover, we have extended our analysis to establish a connection between particle creation rates in the vacuum and those in a medium.
This extension has facilitated an understanding of how the number density of produced particles evolves during the preheating process.

Our approach can address both parametric resonance and tachyonic instability within the same framework. 
In the case of tachyonic instability, we have explored two models. One is tachyonic preheating, 
in which the inflaton possesses a trilinear coupling with another scalar and can have both tachyonic instability and broad resonance effects.
The other model is gauge preheating, where the 
inflaton interacts with gauge bosons via Chern-Simons terms. 
Notably, all our analytic findings are consistent with prior literature.

Crucially, the effective action approach offers an alternative perspective on the preheating process and serves as a valuable complement to
the well-established WKB method. The WKB computation relies on satisfying adiabatic conditions.
When inflaton crosses its potential minimum, the adiabatic condition is violated. Then, we solve the mode functions
before and after passing the minimum using the WKB method, connect the solutions by analytic continuation.
On the other hand, the effective action method determines semiclassical solutions near the potential's minimum,
where particle production predominantly occurs.
This distinction clarifies that these two methods cover different phases of the inflaton oscillation period. 
It also explains that the effective action method does not 
capture the phase accumulating during the oscillating period.

Our work stresses that effective action is a powerful tool for investigating non-perturbative particle production processes. 
Its applicability extends to various domains within cosmology and gravity, opening up alternative ways for computation and 
offering new perspectives on these subjects.
As we look to the future, there is significant potential for applying this approach to scenarios such as 
other well-motivated preheating models, particle production near
black holes, in an expanding universe, or during bubble nucleation.







\acknowledgments

We are grateful to Sergey Sibiryakov 
for useful discussions. 
This work was supported in part by the United States Department of Energy
under Grant No. DE-SC0022148.

\appendix

\section{Effective Action for a Periodic Potential}

In this appendix, we provide the calculation of the effective action for a periodic potential.
Following the preheating model given in \cref{sec:preheating} of parametric resonance, 
the effective action for each mode is written as   
\begin{equation}\label{sk}
	S_\mathbf{k} =- \frac{i}{2}  \int_0^\infty \frac{ds}{s} e^{-s(\mathbf{k}^2+m^2)}
	\int {\rm d}\tau   \langle \tau|e^{-\hat{H}s}|\tau\rangle
\end{equation}
where we have rotated both time $t$ and Schwinger time $s$ into Euclidean coordinates: 
$t \to i \tau$ and $ s \to - i s$, and $\hat{H}=p^2-g^2\phi_0^2\sinh^2(M_\phi\tau)$.

\subsection{A cut Fock Space}

We first introduce the concept of the cut Fock space, which enables us to evaluate the eigenvalues of $\hat{H}$ numerically. 
The time integration is then converted into a summation of eigenstates of $\hat{H}$.
\begin{equation}
	\int {\rm d}\tau \langle \tau| {\rm e}^{- \hat{H} s }| \tau \rangle=\sum_n e^{-s E_n}  \, ,
\end{equation}
Here $\{|n\rangle\}$ are the eigenstates of the harmonic oscillator with Hamiltonian $\hat H_0=p^2+\omega^2\tau^2$. 
The ladder operators $a,a^\dagger$ act as follows,
\begin{equation}
	a|n\rangle=\sqrt{n}|n-1\rangle,\quad a^\dagger|n\rangle=\sqrt{n+1} |n+1\rangle
\end{equation}
resulting in their matrix representations,
\begin{equation}
	a_{kj}=\delta_{k,j-1}\sqrt{k},\quad a^\dagger_{kj}=\delta_{k-1,j}\sqrt{k-1} \, .
\end{equation}
Consequently, the coordinate and momentum operators can be expressed as,
\begin{align}
	\tau_{kj}&=\frac{1}{\sqrt{2\omega}}(a^\dagger+a)_{kj}=\frac{1}{\sqrt{2\omega}} (\delta_{k-1,j}\sqrt{k-1}+\delta_{k,j-1}\sqrt{k})\\ 
	p_{kj}&=i\sqrt{\frac{\omega}{2}}(a^\dagger-a)_{kj}=i\sqrt{\frac{\omega}{2}} (\delta_{k-1,j}\sqrt{k-1}-\delta_{k,j-1}\sqrt{k})
\end{align}
This allows us to diagonalize the Hamiltonian,
\begin{equation}
	\hat H_{0,kj}=2\omega(a^\dagger a+\frac{1}{2})_{kj}=2\omega(k+\frac{1}{2})\delta_{k,j}
\end{equation}
By limiting the basis to $n\le N$, we truncate the matrix representation to a finite $(N+1)\times(N+1)$ dimensional matrix space. 
We can then numerically solve the Hamiltonian by diagonalizing the matrix in its $(N+1)\times(N+1)$ form to obtain its spectrum and eigenstates.

For the inverted harmonic oscillator, we perform an analytic continuation of $\omega\to -i\omega$, resulting in
\begin{equation}
	\hat H'_0=p^2+(-i\omega)^2\tau^2=p^2-\omega^2\tau^2=-2i\omega(k+\frac{1}{2})\delta_{k,j}
\end{equation}
The eigenvalues $E_n=-2i\omega(k+\frac{1}{2})$ have negative imaginary parts, indicating that the probability of the eigenstate $|\psi_n\rangle$ 
decays exponentially,
\begin{equation}
	P_n(s)=|\langle\psi_n|e^{-is\hat H'_0}|\psi_n\rangle|^2=e^{2\mathrm{Im}[E_n]s}=e^{-4\omega(n+\frac{1}{2})s}
\end{equation}
Since the potential of the Hamiltonian $\hat H'_0$ is not bounded from below, any state will infinitely decay downwards.

Now, when we apply a periodic potential, we perform a similar analytic continuation
\begin{equation}
	\hat H'_1=p^2+g^2\phi_0^2\sin^2(iM\tau)=p^2-g^2\phi_0^2\sinh^2(M\tau) \, ,
\end{equation}
which is also unbounded from below. 
We find that some of the eigenvalues of the truncated Hamiltonian $\hat H'_1$ have positive imaginary parts, indicating 
an exponential increase in probability.
However, these eigenvalues are not physical and arise from the truncation of the Fock space to a finite dimension.
Thus, we remove them in the numerical calculations.

When performing the Schwinger time integration, we need the regulator to remove the analytic singularity, defined as 
\begin{align}\label{reg}
	\int d\tau e^{s g^2\phi_0^2\sinh^2M_\phi\tau}\int \frac{dp}{2\pi} e^{-sp^2} = \frac{1}{M_\phi\sqrt{4\pi s}} e^{-g^2\phi_0^2s/2} K_0(-g^2\phi_0^2s/2) 
   \, .
\end{align}
Here, $K_0(z)$ is the zeroth-order modified Bessel function of the second kind \cite{abromowitz1972handbook}.
After applying this regulator, there is still a numerical singularity near $s=0$ when using a partial sum up to the $N$-th energy level of 
the Hamiltonian,
\begin{equation}
	\sum_{n=0}^N e^{-is E_n} \xrightarrow{s\to 0} N
\end{equation}
To resolve this, we truncate at a small value of $s$, $s_{cut}$, and extrapolate to $s=0$ using the second-order extrapolation $f_{ex}(s)$:
\begin{equation}
    f_{ex}(s)=f(s_{cut})+f'(s_{cut})(s-s_{cut})+\frac{f''(s_{cut})}{2}(s-s_{cut})^2 \, ,
\end{equation}
where
\begin{equation}
    f(s)=\sum_{n=0}^N e^{-is E_n}  \, .
\end{equation}
We have discussed the cut fock space method in the context of a periodic potential. Now, Let us compare it to 
the case of a quadratic potential for which we have analytic results. The parameter settings are as follows
\begin{equation}
	g\phi_0=10,\quad M_\phi=0.1,\quad \mathbf{k}^2+m^2_\chi=1 \, ,
\end{equation}
For the inverted harmonic oscillator, the decay rate is calculated as, 
\begin{equation}
	2\mathrm{Im}S_\mathbf{k}=\frac12\ln\left(1+e^{-\frac{\pi(\mathbf{k}^2+m^2_\chi)}{g\phi_0 M_\phi}}\right)=\frac12\ln(1+e^{-\pi})=0.02115 \, .
\end{equation}
Regarding the periodic potential, the numerical method ($N_{\text{cut}}=100,s_{\text{cut}}=0.2$) yields
\begin{align}
	2\mathrm{Im} S_\mathbf{k}=&-\frac12\int_{s_{cut}}^\infty \frac{ds}{s} 
	e^{-s(\mathbf{k}^2+m^2)} \left(
	\sum_{n=0}^{N_{cut}} e^{-s E_n}-\frac{1}{M_\phi\sqrt{4\pi s}} e^{-g^2\phi_0^2s/2} K_0(-g^2\phi_0^2s/2) \right)\nonumber\\
	&-\frac12\int_0^{s_{cut}}f_{ex}(s)ds=0.02212\pm 0.00002 \, .
\end{align}
The two results are consistent.

\subsection{Instanton method}
In this section, we introduce another method for calculating the effective action by finding the instanton solution for a periodic potential. 
We express the expectation value of the evolution operator as
\begin{equation}
	\langle \tau |e^{-\hat{H}s}| \tau \rangle \simeq e^{-S_0}\det\left(-\frac{1}{2}\frac{d^2}{ds^2}+V''(\tau ) \right)^{-\frac{1}{2}}
\end{equation}
where $V=-g^2\phi_0^2\sinh^2(M\tau)$.
The Lagrangian of the system is given as $L=\frac{1}{4}\dot{\tau}^2-g^2\phi_0^2\sinh^2(M\tau)$, and the energy
$C=\frac{1}{4}\dot{\tau}^2+g^2\phi_0^2\sinh^2(M\tau)$ is a constant of motion. It gives us the solution
\begin{equation}
	2iM\sqrt{C}(s+s_0)=F(iM\tau|-\frac{g^2\phi_0^2}{C}) \, ,
\end{equation}
indicating that
\begin{equation}
	\sin(iM\tau)=\text{sn}(2iM\sqrt{C}(s+s_0)|-\frac{g^2\phi_0^2}{C}) \, , 
\end{equation}
or equivalently,
\begin{equation}
	\sinh(M\tau)=\sqrt{\frac{C}{C+g^2\phi_0^2}}\text{sd}(2M\sqrt{C+g^2\phi_0^2}(s+s_0)|\frac{C}{C+g^2\phi_0^2})
\end{equation}
where $F(\varphi|m)$ is the elliptic integral of the first kind, $\text{sn}(u|m)$ and $\text{sd}(u|m)$ are Jacobi elliptic functions 
\cite{abromowitz1972handbook}. 
To simplify the notation, we define
\begin{equation}
	\varphi\equiv M\tau,\quad m\equiv\frac{C}{C+g^2\phi_0^2},\quad u\equiv 2M\sqrt{C+g^2\phi_0^2}s=\frac{2g\phi_0 M}{\sqrt{1-m}}s \, .
\end{equation}
The solution is now expressed as 
\begin{equation}
	\sinh\varphi=\sqrt{m}\text{sd}(u+u_0|m) \, .
\end{equation}
The periodic boundary condition $\tau(0)=\tau(\bar s)=\tau_0$ (or $\varphi(0)=\varphi(\bar u)=\varphi_0$) implies that
\begin{equation}
	\bar u+2u_0=(4n+2)K(m) \text{, and } \sinh\varphi_0=\sqrt{m}\text{sd}(u_0|m)
\end{equation}
By replacing $u_0=(2n+1)K-\frac{\bar u}{2}$, we finally arrive at the form
\begin{equation}
	\sinh\varphi=\sqrt{\frac{m}{1-m}}\text{cn}(u-\frac{\bar u}{2}|m) \, ,
\end{equation}
or equivalently,
\begin{equation}
	\cosh\varphi=\frac{1}{\sqrt{1-m}}\text{dn}(u-\frac{\bar u}{2}|m)\text{, or } \tanh\varphi=\sqrt{m}\text{cd}(u-\frac{\bar u}{2}|m)
\end{equation}
where $m(\tau_0,\bar s)$ is determined by
\begin{equation}\label{eqm}
	\sinh M\tau_0=\sqrt{\frac{m}{1-m}}\text{cn}(\frac{g\phi_0 M\bar s}{\sqrt{1-m}},|m)
\end{equation}
By using the following expressions,
\begin{equation}
	\frac{du}{ds}=\frac{2g\phi_0 M}{\sqrt{1-m}},\quad \dot\tau=\frac{2g\phi_0}{\sqrt{1-m}}\frac{d\varphi}{du},\quad  \frac{d\varphi}{du}=-\sqrt{m}\text{sn}(u-\frac{\bar u}{2}|m)
\end{equation}
the Lagrangian can now be written as
\begin{equation}
	L=\frac{g^2\phi_0^2}{1-m}\left(\frac{d\varphi}{du}\right)^2-g^2\phi_0^2\sinh^2(\varphi) = \frac{g^2\phi_0^2m}{1-m}(\text{sn}^2-\text{cn}^2)=\frac{g^2\phi_0^2m}{1-m}(2\text{sn}^2-1)  \, ,
\end{equation}
and the classical action takes the form as
\begin{align}
	S_0&=\int_0^s ds' L=\int_0^{\bar u} \frac{\sqrt{1-m}}{2g\phi_0 M}du \frac{g^2\phi_0^2m}{1-m}(2\text{sn}^2-1) \\
	&= \frac{g\phi_0 m}{M\sqrt{1-m}}\left[Sn(\frac{\bar u}{2})-Sn(-\frac{\bar u}{2})-\frac{\bar u}{2}\right]\\
	&=\frac{g\phi_0}{M\sqrt{1-m}}\left[-2E(\frac{\bar u}{2})+(1-\frac{m}{2})\bar u\right]
\end{align}
Considering the limit that $m\to 0 (H\to 0)$:
\begin{equation}
	S_0\approx-\frac{g\phi_0}{M}\varphi^2_0\tan\frac{\bar u}{2}=S_0^{SHO} 
\end{equation}
implying that the instanton can be approximated as a harmonic oscillator when the energy is small. 
Other limiting cases are
\begin{gather}
	S_0\approx\frac{\pi n g \phi_0}{8M}m^2\to 0 \text{, for } m=\tanh^2\varphi_0\to 0  \, , \\
	S_0\approx\frac{n g\phi_0}{M\sqrt{1-m}}(\ln\frac{16}{1-m}-4)\to\infty \text{, for } m=\tanh^2\varphi_0\to 1 \, .
\end{gather}

Now, let us evaluate the determinant with the Gelfand-Yaglom formula. One can check that
\begin{equation}
	\tau_1=\text{sn}(u-\frac{\bar u}{2}) \propto \dot \tau_{cl}
\end{equation}
satisfies the eigenvalue equation
\begin{equation}
	\frac{\delta^2 S}{\delta\tau^2}\tau=\left(-\frac{1}{2}\frac{d^2}{ds^2}+V''\right)\tau=0
\end{equation}
So we can use the Wronskian $W=\tau_1\dot\tau_2-\tau_2\dot\tau_1$ to find $\tau_2$. 
By using $\dot W=0$ and
\begin{equation}
	\frac{d}{ds}\left(\frac{\tau_2}{\tau_1}\right)=\frac{W}{\tau_1^2}
\end{equation}
we obtain that
\begin{equation}
	\tau_2=W\tau_1\int_0^s\frac{ds'}{\tau_1^2}, \dot\tau_2=W\left(\dot\tau_1\int_0^s\frac{ds'}{\tau_1^2}+\frac{1}{\tau_1}\right)
\end{equation}
Thus, we have
\begin{equation}
	\tau_2(0)=0, \dot\tau_2(0)=\frac{W}{\tau_1(0)}
\end{equation}
By taking $W=\tau_1(0)$, we get 
\begin{align}
	\tau(s)&=\tau_1(0)\tau_1(s)\int_0^s\frac{ds'}{\tau_1^2(s')}=-\text{sn}(\frac{\bar u}{2})\text{sn}(u-\frac{\bar u}{2})\int_0^u\frac{du'/\dot u}{\text{sn}^2(u'-\frac{\bar u}{2})}
    \nonumber  \\
	&=-\frac{1}{\dot u}\text{sn}(\frac{\bar u}{2})\text{sn}(u-\frac{\bar u}{2})\left[Ns(u-\frac{\bar u}{2})-Ns(-\frac{\bar u}{2})\right]
      \nonumber \\
	&=-\frac{\sqrt{1-m}}{2g\phi_0 M}\text{sn}(\frac{\bar u}{2})\text{sn}(u-\frac{\bar u}{2}) 
         \nonumber \\
      & \times \left[u-E(u-\frac{\bar u}{2})-E(\frac{\bar u}{2})-\text{cn}(u-\frac{\bar u}{2})\text{ds}(u-\frac{\bar u}{2})-\text{cn}(\frac{\bar u}{2})\text{ds}(\frac{\bar u}{2})\right]
\end{align}
Therefore, the determinant is calculated as 
\begin{equation}
	\det\left(-\frac{1}{2}\frac{d^2}{ds^2}+V''\right)=4\pi\tau(\bar u)=-\frac{4\pi\sqrt{1-m}}{g\phi_0 M} \text{sn}^2(\frac{\bar u}{2})\left[\frac{\bar u}{2}-E(\frac{\bar u}{2})-\text{cn}(\frac{\bar u}{2})\text{ds}(\frac{\bar u}{2})\right]
\end{equation}
The evolution operator 
\begin{equation}
	\langle \tau|e^{-s \hat H}|\tau\rangle=\frac{1}{\sqrt{det(m)}}e^{-S_0(m)}
\end{equation}
is a parametric function of $\tau_0$ and $\bar s$ with $m$ determined by \eqref{eqm}. 
One can change the variable by 
\begin{equation}
	\frac{\partial\tau_0}{\partial m}=\frac{\left(g\phi_0 M \bar s-\sqrt{1-m} E\left(\frac{g\phi_0 M \bar s}{\sqrt{1-m}}\right)\right) \text{nd}\left(\frac{g\phi_0 M \bar s}{\sqrt{1-m}}\right) \text{sd}\left(\frac{g\phi_0 M \bar s}{\sqrt{1-m}}\right)-\sqrt{1-m} \text{cd}\left(\frac{g\phi_0 M \bar s}{\sqrt{1-m}}\right)}{2 \sqrt{(1-m) m} M \left(m \text{cd}\left(\frac{g\phi_0 M \bar s}{\sqrt{1-m}}\right)^2-1\right)}
\end{equation}
Combing the above results, we deduce the effective action, given as
\begin{equation}
	S_{\bf k}=-\frac{1}{2}\int \frac{ds}{s} e^{-s(\mathbf{k}^2+m_\chi^2)} \int_0^1 dm \left|\frac{\partial\tau_0}{\partial m}\right| \frac{1}{\sqrt{det(m)}}e^{-S_0(m)} 
\end{equation}
We did not find a final analytic result of integration, but we can evaluate it numerically.
Also, it could be approximated using the steepest descent method near $\varphi=0$. We then obtain the same result as the quadratic potential.

\bibliographystyle{JHEP}
\bibliography{ref}

\end{document}